\documentclass[aps,pra,superscriptaddress,showpacs,amsmath,amssymb,amsfonts,reprint,a4paper]{revtex4-1}

\usepackage{hyperref}
\usepackage{graphicx}
\usepackage{amsmath}
\usepackage{bbm}
\newcommand{\be}{\begin{equation}}
\newcommand{\ee}{\end{equation}}
\newcommand{\ben}{\begin{equation*}}
\newcommand{\een}{\end{equation*}}

\newcommand{\tr}{\mbox{Tr}}                                        
\newcommand{\bra}[1]{\ensuremath{\langle #1 |}}
\newcommand{\ket}[1]{\ensuremath{| #1 \rangle}}
\newcommand{\prj}[1]{\ensuremath{| #1 \rangle \langle #1 |}}
\newcommand{\ovl}[2]{\ensuremath{\langle #1 | #2 \rangle}}
\newcommand{\matel}[3]{\ensuremath{\langle #1 | #2 | #3 \rangle}}
\newcommand{\avg}[2]{\ensuremath{\langle{#1}\rangle_{#2}}}

\begin{document}

\title{Average entanglement dynamics in open two-qubit systems with continuous monitoring}

\author{Ivonne Guevara}
\affiliation{Departamento de F\'{\i}sica, Universidad Nacional de Colombia,
  Carrera 30 No.45-03, Bogot\'a D.C., Colombia}
\affiliation{Centre for Quantum Computation and Communication Technology (Australian Research Council), Centre for Quantum Dynamics, Griffith University, Brisbane, Queensland 4111, Australia}
\author{Carlos Viviescas}
\affiliation{Departamento de F\'{\i}sica, Universidad Nacional de Colombia,
  Carrera 30 No.45-03, Bogot\'a D.C., Colombia}
  
\date{\today}

\begin{abstract}
We present a comprehensive implementation of the quantum trajectory theory for the description of the entanglement dynamics in a Markovian open quantum system made of two qubits. We introduce the average concurrence to characterize the entanglement in the system and derive a deterministic evolution equation for it that depends on the ways information is read from the environment. This builded in flexibility of the method is used to address two actual issues in quantum information: entanglement protection and entanglement estimation. We identify general physical situations in which an entanglement protection protocol based on local monitoring of the environment can be implemented. Additionally, we methodically find unravelings of the system dynamics providing analytical tight bounds for the unmonitored entanglement in the system at all times. We conclude by showing the independence of the method on the choice of entanglement measure. 
\end{abstract}
\maketitle

\section{Introduction}

At the core of the Physics behind quantum information theory lays the fundamental fact that different constituents of a quantum system can interfere coherently among themselves, giving rise to correlations that are absent in classical systems. Prospects of technological applications of this theory, from quantum communication to quantum computation \cite{Niel10}, relay ultimately on our capacity to harness these correlations under physically realistic conditions---that is, in the deleterious presence of the decoherence induced by the unavoidable interaction of the quantum systems with their surroundings---, a goal, that despite the enormous effort invested during the last years in its pursuit, has been only partially achieved. The fragility of quantum correlations to perturbations, a shortcoming which is magnified in the face of the transition from microscopic to macroscopic scales demanded for applications, persists arguably as one of the main hurdles for the emerging quantum technologies and a topic of which our knowledge remains limited. Until recently \cite{Konr07,Tier08,Gour10}, for instance, our understanding of the time evolution of entanglement---a prominent embodiment of these quantum correlations, and a recognized key resource for quantum information processing and communication \cite{Horo09}---under the effects of decoherence had barely increased beyond what we had gathered during its initial exploration period, as the shortage of general results both theoretical and experimental can testify \cite{Zyczkowski:2001fs,Roos:2004hm,Dodd:2004iw,Carv04,Dur:2004cx,Yu:2006ha,Fine:2005iz,Santos:2006fo,Almeida:2007fb,Terr07}.

The essential difficulty faced by any attempt of formulating a complete description of the entanglement dynamics in open quantum systems resides on the nonlinear dependence of entanglement with the system state. Not surprisingly, therefore, the traditional head-on approach to the problem, in which the time evolution of the system state is first solved and only afterwards, for each point in time, the state entanglement is estimated, yields limited results. The necessary resources this strategy demands, both computationally and experimentally, pile up very rapidly with the system size, strongly restricting its application \cite{Zyczkowski:2001fs,Roos:2004hm,Dodd:2004iw,Carv04,Dur:2004cx,Yu:2006ha,Fine:2005iz,Santos:2006fo,Almeida:2007fb,Terr07}. In recent years, however, new insight on the subject was gained from exploiting symmetries of the entanglement measures used, which led to the formulation of an efficient dynamical equations for entanglement in composite systems in which a single one of its constituents is coupled to a noisy channel \cite{Konr07,Tier08,Gour10}, and opened a path to further generalizations \cite{Gheo12}. 

Fruitful results have also emerged lately from a different perspective altogether on the problem. In the quantum trajectory description of open systems \cite{Carm93} the state of the system can be resolved at all times if information is being read from the environment in a continuous way, hence the state evolution is no longer deterministic but stochastic, and conditioned to the measurement record of the environment. As a consequence, also the entanglement in the system vary according to how the environment is being monitored, and its time evolution can then be systematically characterized in an efficient way \cite{Nha04,Carv07}. Numerical and analytical explorations of this proposal allowed for the description of the dynamics of  entanglement \cite{Nha04,Carv07,Vivi10,Voge10} and entanglement of assistance \cite{Masc11} in various experimentally relevant situations, and preluded the formulation of new protocols for entanglement protection \cite{Carv11} and quantum computation \cite{Santos:2012ed}, as well as its extension for the study of general quantum correlations \cite{Grim12}. Here we present a rigorous implementation of the quantum trajectory theory for the systematic study of the dynamics of entanglement in open quantum systems of two-qubits, and lay with clarity the groundwork for its extension to quantum systems of arbitrary size. The strength of our operational method is demonstrated with a methodic approach to the topics of entanglement protection and analytical bound estimation for the entanglement dynamics of the unmonitored system state.

Central to our description of the entanglement dynamics in $2 \times 2$ Markovian open systems is the issue of its quantification in the emerging ensemble of quantum trajectories. We address this problem in Section~\ref{sec:QTE} using the formalism introduced in Section~\ref{sec:QT}. For this we adopt ideas put forward in \cite{Nha04,Carv07} and use the \emph{average} concurrence as entanglement measure---a choice motivated only by convenience, as it will become apparent---. The quantum trajectory theory then allows for a methodical derivation of a \emph{deterministic} equation for the dynamical evolution of the average concurrence, whose solution does not require the complete knowledge of the state of the system at all times.  Moreover, concomitant with the framework of quantum trajectory theory, the dynamical portrait of entanglement provided by this equation of motion changes depending on the measurement scheme chosen to monitor the environment, granting the method a versatility and control that can potentially be exploited for quantum information related tasks. We demonstrate this point by exhibiting two different facets of our formalism: In combination with only local environments, in Section~\ref{sec:local} we propose local measurement strategies of the environments to effectively and efficiently protect entanglement in the system, generalizing concepts presented in \cite{Carv11}. In contrast, considering non local monitoring schemes, in Section~\ref{sec:RhoE} we develop computational methods to obtain tight dynamical bounds on the entanglement of the unconditional state of the system, which outperform previously reported ones \cite{Konr07,Tier09,Gour10}. To conclude, in Section~\ref{sec:EoF} we discuss the implementation of the method for entanglement of formation as an example of the use of measures different to concurrence.


\section{\label{sec:QT}Continuous Markovian Unravellings}

We study the entanglement dynamics in open quantum systems which are weakly coupled to large environments and follow a Markovian dynamics dictated by an evolution equation of the Lindblad form \cite{Lind76}
\be
\label{eq:MEq}
\dot\rho= -\frac{i}{\hbar} [H,\rho] - \sum_{k=1}^{L} \mathcal{D}[J_k] \rho  \,.
\ee
Here, the unitary dynamics generated by the system Hamiltonian $H$ is accounted in the first term on the right-hand side, while the effects of the environment on the system state $\rho$ are included through 
\ben
\mathcal{D}[J_{k}]\rho \equiv J_k \rho J_k  - \frac{1}{2} ( J_k^{\dagger}J_k\rho + \rho J_k^{\dagger}J_k) \,,
\een
with Lindblad operators $J_k$ ---which we will lump together as a vector $\mathbf{J}=(J_1,\dots,J_L)^{\mathsf{T}}$ hereafter--- fixed by the specific coupling between system and environment.

The separation of times scales implied in the above master equation, means that for this kind of systems the environment can be continuously measured on times much shorter than any characteristic times of the system. Assuming such a continuous monitoring scheme of the environment is implemented with perfect efficiency, the measurement record obtained in this way yields information about the system. Ignore this record, and the disregarded information implies that a pure state of the system is transformed into a statistical mixture \cite{vonN96,Krau83} with a time evolution given by master equation \eqref{eq:MEq}. If on the other hand, one does take note of the outcomes of the measurements, then the state of the system immediately after the measurement is again pure, changing stochastically and  conditioned on the measurement outcomes. In the later case the evolution of the system, induced by the monitoring, describes a \emph{quantum trajectory} on the system Hilbert space  \cite{Carm93,Wise96,Perc98,Rigo97,Wise01}. Both procedures reconcile by noticing that at a given time the system unconditional state $\rho$ is recovered upon averaging the conditioned state $\ket{\psi_c}$ over an ensemble of independent realizations of the stochastic quantum trajectories,
\ben
\label{eq:rhodeco}
\rho(t) = E[\ket{\psi_c (t)}\bra{\psi_c (t)}]
\een
with $E[\cdot]$ denoting the ensemble average. Since in this manner at each instant of time a pure state decomposition of the density matrix of the system is obtained, we say that the quantum trajectories \emph{unravel} the master equation. The existence of a plurality of unravelings is a remarkable feature of this approach which is in correspondence to the many different ways in which the environment can be continuously measured \cite{Carm93} and lies at the heart of our work.

In this paper we concentrate on the diffusive unravellings \cite{Perc98,Rigo97,Wise01,Wise09}, however our methods can be implemented for all other kind of unravelings. In its It\^o form, the nonlinear stochastic equation representing a general diffusive unraveling and determining the evolution of the conditional state $\ket{\psi_c}$ is
\be
\label{eq:diffeq}
\ket{d\psi_c} = \ket{v(\psi_c)} dt + d\boldsymbol{\xi}^{\dagger}\ket{\mathbf{f}(\psi_c)}  \,.
\ee
The explicit form of the drift and diffusive amplitudes is, respectively,
\begin{subequations}
\begin{align}
\label{eq:drift}
\ket{v(\psi_c)} &= \left[-\frac{i}{\hbar}H -\frac{1}{2} \left(\mathbf{J}^{\dagger} \mathbf{J}
      + \langle \mathbf{J}^{\dagger}\rangle_{c} \langle \mathbf{J}
      \rangle_{c} -2 \langle \mathbf{J}^{\dagger}\rangle_{c}
      \mathbf{J} \right)\right] \ket{\psi_c} \,, \\
\label{eq:diff}
\ket{\mathbf{f}(\psi_c)} &=  (\mathbf{J} - \langle \mathbf{J} \rangle_{c} )\ket{\psi_c}\,,      
\end{align}
\end{subequations}
where the expectation values $\avg{O}{c}$ for any operator $O$ are taken with respect to the conditional state $\ket{\psi_c}$, and $\dagger$ represents transpose $(\mathsf{T})$ of the vector and Hermitian adjoint of its components. The stochastic nature of the time evolution is incorporated in the noise term through the vector $d\boldsymbol{\xi}=(d\xi_1,\dots,d\xi_L)^{\mathsf{T}}$ composed of 
infinitesimal complex Wiener increments \cite{Gard85}. The random process $d\boldsymbol{\xi}$ has vanishing  ensemble average,
$E[d\boldsymbol{\xi}]=0$,  
with correlations 
\begin{equation}
\label{eq:dxi}
d\boldsymbol{\xi}d\boldsymbol{\xi}^{\dagger}= \mathbb{I}\,dt\,, \quad
d\boldsymbol{\xi}d\boldsymbol{\xi}^{\mathsf{T}}= u\, dt\; ,
\end{equation}
where $\mathbb{I}$ is the identity matrix and $u$ is a $L\times L$ complex symmetric matrix. Physical choices of $u$ are restricted by the condition that the $2L\times 2L$ correlation matrix $R$ for the real vector $(\text{Re}\,d\boldsymbol{\xi}, \text{Im}\,d\boldsymbol{\xi})^{\mathsf{T}}$,
\be
\label{eq:R}
R \equiv \frac{dt}{2} \begin{pmatrix}
\mathbb{I} + \text{Re}\,u & \text{Im}\, u \\
\text{Im}\, u & \mathbb{I}- \text{Re}\,u
\end{pmatrix}\,,
\ee
is positive definite. 

Associated to each unraveling is the measurement record upon which the evolution of $\ket{\psi_c}$ is conditioned. It can be written as the vector of complex currents \cite{Wise01}
\begin{equation}
\label{eq:current}
  \mathbf{Y}^{\mathsf{T}}\, dt= \langle \mathbf{J}^{\dagger} u  +
    \mathbf{J}^{\mathsf{T}} \rangle_{c} \, dt + d\boldsymbol{\xi}^{{\mathsf{T}}}\,,
\end{equation}
where each component represents a specific detection event. A diffusive unraveling is then completely specified once the correlation matrix $u$ is given and the experimental setup for measuring the environment fixed. 

Although all unravelings of the master equation are equivalent, i.e., they all lead on average to the same unconditional solution $\rho$ of the master equation, we notice that the access to different unravelings of the master equation is achieved by a proper tuning of the noise correlations. Different choices of $u$ lead to ensembles of quantum trajectories with distinct statistical properties, and, consequently, to pure state decompositions of the unconditional state with varying statistical features. Such differences are irrelevant for the evaluation of expectation values of linear operators because $\avg{A}{} = \tr(\rho A) $ for all linear operators $A$, yet, as we show below, for the quantification of entanglement in the ensemble of quantum trajectories they are of the utmost significance.

\section{\label{sec:QTE}Quantum trajectory entanglement dynamics}

The use of quantum trajectory theory to describe the time evolution of an open quantum system offers new paths for the characterization of the entanglement in an ensemble of stochastic trajectories \cite{Nha04,Carv07} that goes beyond the standard approach to the entanglement dynamics. In order to offer a clear picture of these new ideas, in the remaining of this paper we concentrate on two-qubit systems 
coupled to different kind of environments, and study their impact on the entanglement of the system. In all cases we take a pure state as the initial state of the dynamics, and, since our primarily interest is on the decoherence effects, we assume the qubits do not interact directly, i.e., the system Hamiltonian only induces local transformations. For the quantification of the entanglement in the system we chose concurrence, a widespread and well studied entanglement measure, which is easily computable for the systems at hand \cite{Woot98}. These choices are made for the sake of concreteness only, as our arguments and conclusions can be extended straightaway to composite systems of arbitrary dimensions \cite{Vivi14} and to the use of other entanglement measures, e.g., entanglement of formation \cite{Woot98} (see Section~\ref{sec:EoF}) or SL-invariant measures \cite{Vers03,Gour10,Vivi14b}.

\subsection{Average concurrence}

Within the quantum trajectory formalism the entanglement associated to a particular unraveling $u$ of the open system dynamics is accounted in a natural way \cite{Nha04,Carv07}. In as much as along a quantum trajectory the conditional state of the system remains pure, we utilize concurrence to quantify its entanglement. The concurrence $C(\psi)$ of a general pure state of a two-qubit system $\ket{\psi}=\psi_{00}\ket{00}+\psi_{01}\ket{01}+\psi_{10}\ket{10}+\psi_{11}\ket{11}$ is defined as the absolute value of the preconcurrence $c(\psi)\equiv \ovl{\psi}{\tilde{\psi}}=2(\psi_{01}^*\psi_{10}^*-\psi_{00}^*\psi_{11}^*)$, i.e., $C(\psi) \equiv |c(\psi)|$ \cite{Woot98}. Here $\ket{\tilde{\psi}} = \sigma_y \otimes \sigma_y \ket{\psi^{*}}$ is the corresponding spin flip state, with $\ket{\psi^{*}}$ the complex conjugate of $\ket{\psi}$ in the computational basis and $\sigma_y$ the second of Pauli matrices. 
Consider now an ensemble of quantum trajectories corresponding to an unraveling $u$ of the master equation. Along each trajectory, following the evolution of the conditional state $\ket{\psi_{c}}$, the entanglement jumps in a random manner giving rise, at each time point $t$, to a distribution of concurrence over the ensemble of conditional states belonging to different quantum trajectories. The entanglement in the ensemble is then naturally characterized by the mean of this distribution,
\be
\label{eq:ECu}
C_{u}(t) \equiv E[C({\psi_{c}}(t))]\,.
\ee

The average concurrence defined in this way depends on the selected unraveling $u$ of the master equation and therefore quantifies the entanglement in the ensemble of quantum trajectories associated to an specific way of monitoring the environment. As a consequence, different monitoring strategies of the environment lead to different entanglement contents in the system, lending a versatility to the method that is unavailable when the entanglement content of the unconditional state is taken as the fundamental quantity characterizing the ensemble entanglement. This, as we show below, can be used to approach a variety of entanglement central issues.

\subsection{Average concurrence dynamics} 

With our choice of the average concurrence $C_u$ to characterize the entanglement in an unraveling of the master equation, we now turn to the description of its time evolution. For a given unraveling $u$ of the master equation, the dynamical equation for the concurrence change $dC(\psi_c)$ along a single quantum trajectory follows from the stochastic evolution equation \eqref{eq:diffeq} for the conditional system state (see Appendix \ref{app:E_cond_change}),
\be
\label{eq:dcon}
dC(\psi_c) = V(\psi_c,u)dt + \text{Re}\left[d\boldsymbol{\xi}^{\dagger}\mathbf{F}(\psi_c) \right].
\ee
The explicit dependence on the unraveling of the open dynamics is displayed on the deterministic term amplitude
\be
\label{eq:V}
\begin{split}
V(\psi_c,u) &= -\text{Re}\left[ \frac{c(\psi_c)}{C(\psi_c)} \left(\avg{\widetilde{\mathbf{J}^{\dagger}\mathbf{J}}}{c} - \frac{1}{c(\psi_c)}|\avg{\widetilde{\mathbf{J}}}{c}|^2  \right.\right. \\
&\quad\left.\left. + \frac{c(\psi_c)}{C(\psi_c)^2} \avg{\widetilde{\mathbf{J}^{\mathsf{T}}}}{c} u^{*} \avg{\widetilde{\mathbf{J}}}{c} - \sum_{kl} \ovl{\widetilde{J_k \psi_c}}{J_l \psi_c} u_{kl}^*\right)\right]
\end{split}
\ee
with the presence of the conjugate matrix $u^*$. As expected, for non interacting particles the effects of the local Hamiltonians on the entanglement evolution vanishes. The noise term amplitude, in contrast, is independent of the unraveling,
\be
\label{eq:F}
\mathbf{F}(\psi_c) = \frac{c(\psi_c) }{C(\psi_c)} \left(\avg{\widetilde{\mathbf{J}}}{c} - c(\psi_c)^{*} \avg{\mathbf{J}}{c} \right) \,.
\ee
To simplify notation, in both equations above we introduced $\avg{\widetilde{O}}{c} \equiv \matel{\tilde{\psi_c}}{O}{\psi_c}$ for any operator $O$.  

The entanglement evolution in the ensemble of trajectories generated by a particular way of gathering information from the environment, as described by $C_{u}(t)$, follows then the dynamical equation 
\be
\label{eq:dcu}
\frac{dC_{u}}{dt} = E[V(\psi_c,u)]\,,
\ee 
obtained after the ensemble average of Eq.~\eqref{eq:dcon}. This evolution law for the average concurrence is the central result of this paper, providing a complete characterization of the entanglement evolution in an open quantum system being continuously monitored.

From the form of the drift term \eqref{eq:V} it is apparent that equation \eqref{eq:dcu} is not, in general, a closed dynamical equation, but its solution entails the integration of a set of coupled equations involving other system observables. The number and nature of these relevant quantities for the determination of the ensemble entanglement evolution depend on the particularities of the system at hand and the environment measurement scheme used. It is remarkable, however, that within our approach they can be identified and a complete closed system of equations constructed whose solution yields the time evolution of the system entanglement.

In the next sections we turn to the application of the previously developed formalism and illustrate its potential to approach different aspects of entanglement dynamics in composite systems. For this we study a collection of simple, yet not trivial, exemplary two-qubit systems coupled to different kind of environments.

\section{\label{sec:local}Local environments and local detection}

In commonly encountered quantum information protocols the parties involved are distant from each other and have access only to their own part of the system, which in turn is coupled to a local reservoir. Information about the system can then be read from the environments only through local measurements. As we now show, for this class of systems Eq.~\eqref{eq:dcu} alone completely determines the time evolution of the system entanglement.

In a system of two noninteracting qubits satisfying the setup just stated the effects of the baths are described by local Lindblad operators of the form $J_k = (J_k\otimes \mathbb{I})$ or $J_k=(\mathbb{I} \otimes J_k)$ acting, respectively, on the first or second qubit. Additionally, within the quantum trajectory approach, local measurements on these independent environments are accounted for by restricting the possible unravelings to $u_{kl} = 0$ if $J_k$ and $J_l$ are operators acting on different qubits, i.e., considering measurement records of local currents only (cf. Eq.~\eqref{eq:current}). Under these conditions Eq.~\eqref{eq:dcu} reduces to (see Appendix~\ref{app:E_cond_change})
\ben
\frac{dC_{u}}{dt} = -k(u)C_{u}\,,
\een 
with the unraveling dependent function
\be
\label{eq:ku}
\begin{split}
k(u) &= \frac{1}{2} \sum_{k} \left[ \tr_{\mathbb{C}^2}(J_k^{\dagger}J_k) - \frac{1}{2}\left| \tr_{\mathbb{C}^2}J_k \right|^2\right] \\ 
&\quad+\frac{1}{2} \text{Re} \left(\sum_{kl} u^{*}_{kl} \left[ \tr_{\mathbb{C}^2}(J_k J_l) - \frac{1}{2}  \tr_{\mathbb{C}^2}J_k   \tr_{\mathbb{C}^2}J_l  \right]\right).
\end{split}
\ee 
This linear dynamical equation can be immediately integrated to obtain
\ben
C_{u}(t) = e^{-\int\limits_{0}^t\!dt' k(u(t'))} C_{0},
\een
which provides a thorough description of the entanglement time evolution in the system when $C_{0}\equiv C(\psi(t=0))$ is the initial state concurrence. 

Hence, for systems coupled to local environments and subject to local measurements the average concurrence dynamics depends solely on the initial state concurrence and is completely determined by the function $k(u)$ alone, which contains the whole information about the unraveling. This clearly indicates that in the search of effective measurement strategies aimed at broad-ranging, i.e., state independent, effects on the entanglement dynamics, e.g., entanglement protection, local monitoring schemes may have the upper hand over other possible approaches. 

In general, local monitoring setups may lead to $k(u)$ which explicitly depends on time, as would be the case if, for example, unravelings with adaptative noise \cite{vanD98,Wise01,Wise09} were considered. Most cases of interest, however, comprise situations in which $k(u)$ is independent of time and, consequently, the entanglement evolution reduces to an exponential dynamics \cite{Voge10},
\be
\label{eq:CuExp}
C_{u}(t) = e^{-k(u)t}C_{0},
\ee 
where the rate $k(u)$ is fixed by the monitoring scheme. 
Such prospect of control over the entanglement evolution in the system  by properly choosing the measurement schemes on the environment widens the spectrum of applications of our formalism, from providing easy to evaluate bounds for the entanglement of the unconditional state (see Section~\ref{sec:RhoE}), to offering local measurement schemes to perfectly shield the system entanglement from the detrimental effects of decoherence, as described in the next section. 

\subsection{\label{Eprotect}Entanglement protection}

The use of quantum trajectories schemes based on local monitoring of the environments for entanglement protection in systems of two-qubits was first proposed  in Ref.~\cite{Voge10} for the specific cases of dephasing and infinite temperature environments. For this last kind of noisy channel the protection scheme was later extended to an arbitrary number of qubits in \cite{Carv11}. In this section we considerably expand this set of examples by using our theory to identify general conditions on two-qubit systems which would allow such protection strategies to be applied.

From the dynamical equation \eqref{eq:CuExp} it is evident that local unravelings yielding $k(u)=0$ fully protect the average concurrence in the system at all times, i.e., $C_u(t)=C_0$ for all $t$. Accordingly, requirements on the environments leading to the aforementioned unravelings can be found by looking at the definition of $k(u)$. A paused inspection of Eq.~\eqref{eq:ku} is enough to convince oneself of the following statement: For systems of two noninteracting qubits, coupled independently to local noisy channels characterized by hermitian Lindblad operators, i.e., $J_k^\dagger = J_k$, local continuous monitoring of the environments corresponding to the unraveling
\be
\label{eq:up}
u^{\text{p}}_{kl}=-\delta_{kl}
\ee  
perfectly protects the entanglement in the system for all initial states; that is, a local continuous measurement of the environments following the prescription given by $u^{\text{p}}$ effectively counteracts the deleterious effects of decoherence on the system entanglement, preserving it despite the system coupling to its surroundings. 

\begin{figure}
	\includegraphics*[scale=.6]{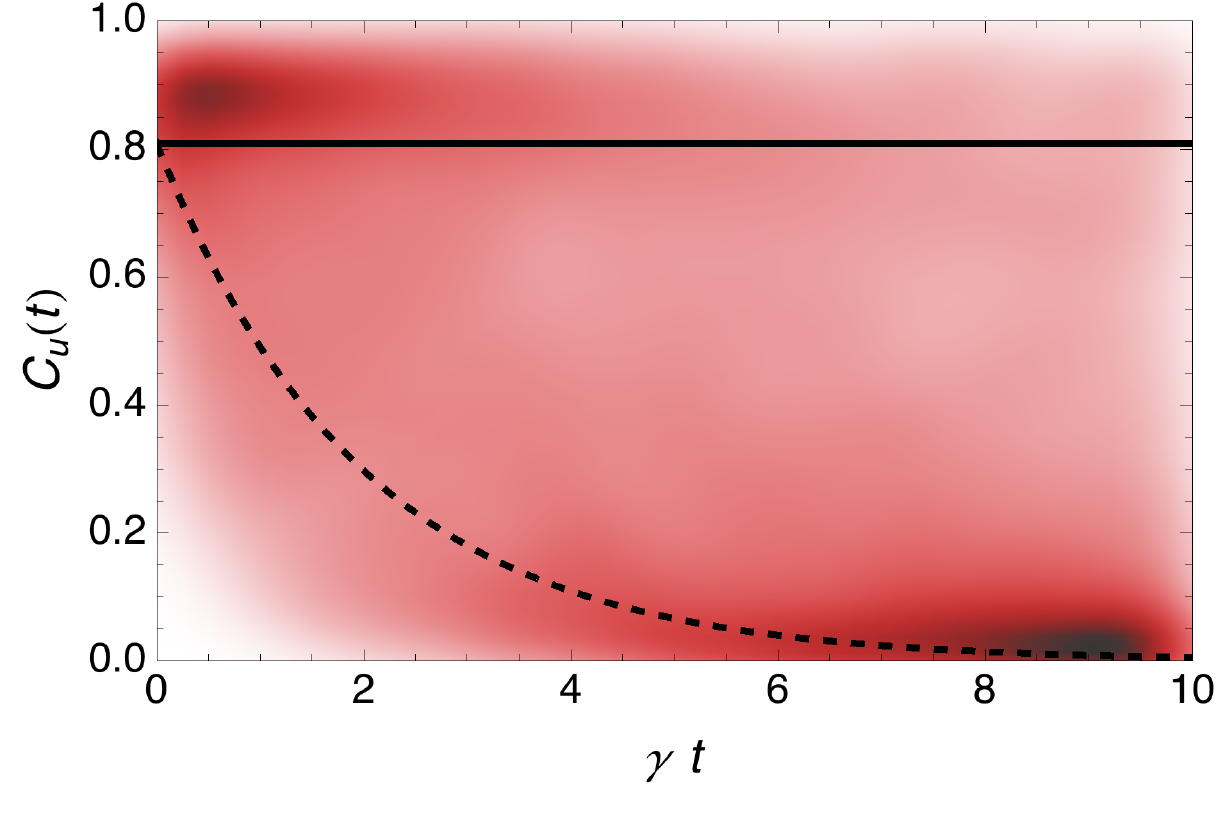}
	\caption{\label{fig:protec} Entanglement protection for a $2\times 2$  system with its first qubit incoherently coupled to a phase damping environment and the second to a thermal bath at infinite temperature. We assume the coupling strength to both baths lead to the same decay rate $\gamma$. Initially the system is prepared in the entangled pure state $\ket{\psi(0)} = (\ket{00} - \ket{01} + i \ket{10} + i\sqrt{5} \ket{11})/\sqrt{8}$ with $C_0=0.809$. If the local unraveling $u^{p}$ is implemented for the continuous monitoring of each of the environments then the entanglement in the system is protected along each \emph{single} quantum trajectory, the concurrence distribution becomes localized and the average concurrence $C_{u^{p}}$ remains constant for all times (solid line). If however,  the protection is only partial, implementing $u^{p}$ for the monitoring of the phase damping channel only, while $u=\begin{pmatrix} -1 & 0 \\ 0 & 1 \end{pmatrix}$ is used for the thermal bath, the average concurrence decays exponentially with $C_{u}(t) = C_0 e^{-2\gamma t}$ (dashed line) but still slower than the unmonitored entanglement. Additionally, concurrence fluctuates randomly on the ensemble of trajectories. At each time the probability density distribution of concurrence is depicted in the background (darker regions correspond to higher densities). For times shorter than the characteristic time of the system entanglement $t\sim 1/2\gamma$ most of the trajectories exhibit an increase in their concurrence, i.e. entanglement creation, a situation that is reversed for larger times when states in most of the trajectories become separable. }
\end{figure}

Even though the argument for the protection protocol was made at the level of the average concurrence, its consequences actually reach deep down to the single trajectory level. Indeed, for the protecting unraveling $u^{\text{p}}$ the stochastic change of the concurrence \eqref{eq:dcon} along a single trajectory reduces to (see Appendix~\ref{app:E_cond_change})
\be
\label{eq:dCp}
dC(\psi_c) =\sum_{k} \text{Re} (d\xi_k^{*}) \left(\frac{1}{2}\tr_{\mathbb{C}^2} J_k  -\avg{J_k}{c} \right)  C(\psi_c),
\ee
where we used that $k(u^{\text{p}})=0$ and $J_k$ is a hermitian operator and hence the expression in parenthesis is real. Notice now that for unraveling \eqref{eq:up} the noise correlations \eqref{eq:dxi} become $d\xi_k d\xi_l^{*} = \delta_{kl}dt$ and $d\xi_k d\xi_l = -\delta_{kl}dt$, and therefore define a purely imaginary noise $d\xi_k=i dW_k$ with real Wiener increments satisfying $dW_k dW_l = \delta_{kl}dt$. As a consequence the right hand side of \eqref{eq:dCp} vanishes and we obtain that along single trajectories belonging to unraveling $u^{\text{p}}$ the system concurrence does not change, $dC(\psi_c) = 0$, but remains equal to the concurrence of the initial state $C_0$, i.e., entanglement is protected. This stronger result, besides providing an obvious explanation to the average concurrence protection, offers insight into the unique dynamics of the ensemble of quantum trajectories of the protecting unraveling: Along any of these trajectories the conditional state of the system jumps from one pure state to another in an stochastic manner, yet all of the states along the trajectory possess the same entanglement as the initial state. Furthermore, since the protection takes place on a single trajectory basis, the entanglement dynamics becomes deterministic and the distribution of entanglement on the whole ensemble of trajectories becomes localized. A portrait of this conclusions is given in Fig.~\ref{fig:protec}.

It is illustrative at this point to look at the currents \eqref{eq:current} for the protecting unraveling, as they provide the record of outcomes conditioning the state evolution. Using definition \eqref{eq:up} for the unraveling they become imaginary,
\ben
Y_k dt = i dW_k,
\een
and consist of purely white noise with average $E[Y_k(t)]=0$. This comes about because $u^{\text{p}}$ selects the measured quadrature of each of the $J_k$ as the one with instantaneous vanishing mean value \cite{Szigeti:2014uf}.

Noisy channels satisfying the required conditions of our entanglement protecting protocol include not only the already alluded dephasing channel, with $J_{\text{dph}} = \sqrt{\gamma}\sigma_{-}\sigma_{+}$, 
and the infinite temperature bath, with $\mathbf{J}_{\infty}=\sqrt{\gamma}(\sigma_{x},\sigma_{y})^{\mathsf{T}}$, but, for example, also the depolarizing channel given by $\mathbf{J}_{\text{dp}}=\sqrt{\gamma}(\sigma_{x},\sigma_{y},\sigma_{z})^{\mathsf{T}}$
\footnote{Throughout the paper $\sigma_{x}$, $\sigma_{y}$, and $\sigma_{z}$ are the Pauli matrices, $\sigma_{+}$ and $\sigma_{-}$ are, respectively, the excitation and deexcitation operators acting on a single qubit, and $\gamma$ the corresponding decay rate for the specific bath being considered.}.
Far from limiting the scope of the protocol, this list of fundamental exemplary baths \cite{Niel10} opens the way for the formulation of more sophisticated strategies for entanglement protection involving, for example, the use of engineered environments, as shown in \cite{Carv11}. Finally, notice that since the protection is of local character not only the two qubits may be coupled to different kind of environments but, depending on the nature of the environments and the available local resources, partial protection can still be achieved, a situation illustrated in Fig.\ref{fig:protec}.

It is worth mention that since our entanglement protecting protocol is deeply rooted on the quantum trajectory theory it is independent of the system size, and can be developed to cover general composite systems with an arbitrary number of parties and dimensions \cite{Vivi14}.

\section{\label{sec:RhoE}Unconditional state entanglement estimation}

\subsection{Mixed state entanglement}

The extension to mixed states of most of the well established entanglement measures for pure states---concurrence included---is built over the pure state decompositions of mixed states: $\rho=\sum_{i} p_i \prj{\psi_{i}}$, with non-negative weights $p_{i}$ and $\sum_{i} p_{i} =1$. For a chosen entanglement measure $\mathcal{E}$ and a given decomposition a sensible approach to the quantification of entanglement would be by the ensemble average $\bar{\mathcal{E}}(\rho) = \sum_{i} p_i \mathcal{E}(\psi_{i})$. However, as it turns out, the way to write a given mixed state as a mixture of (in general nonorthogonal) pure states is not unique, and can be achieved in infinitely many ways. Bona fide entanglement measures for mixed state lift this arbitrariness in different ways, by for example stipulating the entanglement $\mathcal{E}(\rho)$ as the \emph{infimum} over all possible averages $\bar{\mathcal{E}}(\rho)$ \cite{Woot98,Uhlm00,Mint09}, at the cost of introducing a counterintuitive definition whose evaluation very rapidly turns into a daunting numerical problem for systems larger than two qubits. This panorama becomes even somber if a description of the time evolution of entanglement is required, for which almost no general result exist so far \cite{Konr07,Gour10}. Here, with the help of the quantum trajectory approach we face these obstacles and partially overcome them, displaying the capability of the method to render good analytical as well as computationally cheap approximations to the entanglement of the unconditional state in open quantum systems. 
 
To quantify the entanglement of the unmonitored state $\rho$ we select two different entanglement measures for mixed states: The concurrence \cite{Woot98,Uhlm00},     
\be
\label{eq:mixC}
C(\rho) = \inf_{\{p_i, \ket{\psi_i}\}} \sum_{i} p_i C(\psi_i)\,,
\ee
and the concurrence of assistance, 
\be
\label{eq:mixCA}
C_{A}(\rho) = \max_{\{p_i, \ket{\psi_i}\}} \sum_{i} p_i C(\psi_i)\,,
\ee
the latter assess the maximum possible average entanglement that an assistance party can create between the other parties by reducing the state \cite{Divi99,Laus02}. Both measures are uniquely defined through an optimization over all possible decompositions into pure states of the mixed state $\rho$: concurrence by minimizing, and concurrence of assistance by maximizing the mean concurrence of the decompositions. For the case of $2 \times 2$ systems analytical expressions for both expressions exist \cite{Woot98,Laus02}, which allow us to assess the quality of our approximations.

\subsection{Physically realizable pure state ensembles}

One of the main features of the quantum trajectory method is the control it offers over the kind of continuous measurement performed on the environments to yield varying values of $C_u(t)$. We take advantage of this flexibility in order to look for unravelings which generate ensembles of trajectories extremizing the values of the average concurrence, and provide an estimation of the unconditional state concurrence $C(\rho(t))$. For this we proceed from the equation of motion \eqref{eq:dcu}, where it is apparent that a necessary condition for the optimal unraveling $u^{+}$ ($u^{-}$) that minimize (maximize) $C_u(t)$ in a continuous way is that it has to minimize (maximize) the average change $dC_{u}$. We make use of this condition to determine $u^{+}$ ($u^{-}$) by optimizing the parameters which define $u$.

Although at this point the way to approach the quantification of the mixed state entanglement within our method seems natural \cite{Carv07,Vivi10}, its success, as we now explain, is limited by the nature itself of the quantum trajectory ensembles. Notice that in the mixed state entanglement measures \eqref{eq:mixC} and \eqref{eq:mixCA} the optimization is taken over \emph{all} possible decompositions into pure states, regardeless of the way they are generated. The continuous measurement process, however, imposes constrains on the ensembles of states it gives rise to and that represent the unconditional system state at any given time \cite{Wise01b}. These so call \emph{physically realizable} decompositions form then a subset of all possible pure state decompositions in which it is not a priori clear that the optimal decompositions, in the sense of \eqref{eq:mixC} and \eqref{eq:mixCA}, are contained. As a consequence the extent of our results is tempered, and we can only affirm in general that $C(\rho(t)) \le C_{u^{+}}(t)$ and $C_{A}(\rho(t)) \ge C_{u^{-}}(t)$. This notwithstanding, in the following sections we demonstrate in experimentally relevant examples of two-qubit systems that these bounds provide tight approximations---in some cases even exact results---and offer a computable and reliable profile of the unconditional entanglement dynamics in quantum open systems.

\subsection{\label{sec:Dch} Dephasing environments}

As first example we apply the quantum trajectory scheme to the entanglement dynamics of a two-qubit system in which each of the qubits is coupled to its own, independent dephasing channel. Hence, the subsystems do not exchange excitations with their environments, but lose coherence due to elastic scattering. The Lindblad operators describing the effects of the channels on the system are 
\ben
\mathbf{J}= \sqrt{\gamma} (\sigma_{-}\sigma_{+} \otimes \mathbb{I}, \mathbb{I} \otimes \sigma_{-}\sigma_{+})^{\mathsf{T}}\,,
\een
where $\sigma_{+}$ and $\sigma_{-}$ are, respectively, the excitation and deexcitation operators acting on a single qubit, and $\gamma$ the dephasing rate, which we assumed equal for both qubits. Unravelings of the master equation describing the open system dynamics are then specified by a $2\times 2$ matrix $u$. 

After explicit evaluation of Eq.~\eqref{eq:dcu}, the dynamical equation for the average concurrence in the system reads
\be
\label{eq:dCdph}
\begin{split}
\frac{dC_{u}}{dt} &= -\frac{\gamma}{2} C_{u} \\
&\quad - \frac{\gamma}{2} \text{Re} \left( E\left[ \frac{c(\psi_c)}{C(\psi_c)}\left( \psi_{c01}\psi_{c10} (u_{11}^{*} + u_{22}^{*} - 2 u_{12}^{*}) \right.\right.\right. \\
& \quad -  
\left.\psi_{c00}\psi_{c11} (u_{11}^{*} + u_{22}^{*} + 2 u_{12}^{*}) \right) \biggr]\biggr)
\end{split}
\ee
in which different choices of $u$ lead to diverse behaviors for the system entanglement.

For local monitoring of the environment, i.e., $u_{12}=0$, the evolution of the average concurrence becomes exponential (cf. Eq.~\eqref{eq:CuExp}) with rate
\ben
k(u) = \frac{\gamma}{4}\text{Re}(2 + u_{11}^{*} + u_{22}^{*})\,.
\een
Since in this case the dynamics is independent of the initial state, this path offers already two general and simple to evaluate bounds for the system unconditional entanglement. The first choice, $u=\mathbb{I}$, gives the best exponential upper bound for the concurrence evolution of $\rho$,
\ben
C(\rho(t)) \le C_{0} e^{-\gamma t}.
\een
This result coincides with recently proposed bounds for entanglement dynamics \cite{Konr07,Gour10} and its performance has been tested in \cite{Tier09} for short times. For the second choice, $u = - \mathbb{I}$, one gets $k=0$ and entanglement in the system is protected, setting a lower bound for the concurrence of assistance of $\rho$,
\ben
C_A(\rho(t)) \ge C_{0}.
\een
Remarkably, this plain bound already shows that the concurrence of assistance in the system never vanishes. A conclusion that, of course, applies to all systems discussed in Section~\ref{sec:local} too.

Tighter bounds for the unconditional state entanglement are found if non-local measurements of the environment are implemented. Then, after no significant effort, it is verified that the choice of non-local unravelings
\begin{subequations}
\label{eq:Duopt}
\begin{align}
u^{\pm}_{11} = u^{\pm}_{22} &= \pm \frac{1}{2} e^{i \theta_c} \left(e^{i \theta_{a}} - e^{i \theta_{b}} \right), \\
u^{\pm}_{12} &= \mp \frac{1}{2} e^{i \theta_c} \left(e^{i \theta_{a}} + e^{i \theta_{b}} \right),
\end{align}
\end{subequations}
with  $\theta_{c} = \arg (c(\psi_c))$, $\theta_{a} = \arg (\psi_{c01} \psi_{c10})$, $\theta_{b} = \arg (\psi_{c00} \psi_{c11})$, yields extreme values for  \eqref{eq:dCdph}, resulting in the evolution equation
\ben
\frac{dC_{u^{\pm}}}{dt} = -\frac{\gamma}{2} ( C_{u^{\pm}}  \pm E[ X(\psi_c) ])\,
\een
for the average entanglement, where we introduced the conditional state function $X(\psi_c) =2( |\psi_{c01}\psi_{c10}| + |\psi_{c00}\psi_{c11}|)$. Here, a solution to the equation with the plus sign provides an upper bound for the concurrence of the unconditional system state, while a solution of the equation with the minus sign sets a lower bound for the concurrence of assistance of $\rho$. The system of coupled equations is completed with the equation of motion for $E[X(\psi_c)]$ which we evaluated in Appendix~\ref{app:Aux},
\be
\label{eq:AD}
\frac{d}{dt} E[X(\psi_c)] = \mp \frac{\gamma}{2} (C_{u^{\pm}} \pm E[X(\psi_c)]).
\ee
The solution for the average concurrence in these unravelings follows after integration of the above equations, leading to
\be
\label{eq:Cpmdph}
C_{u^{\pm}}(t,\psi_0) = \max\left\{ 0, \frac{1}{2}C_{0}\left(1+e^{-\gamma t}\right) \mp \frac{1}{2}X_0\left(1-e^{-\gamma t}\right)\right\},
\ee
where the maximum is taken since $C_u\ge 0$ is an average over nonnegative quantities and we set $X_0 = X(\psi(0))$. The above result provides analytical, non-trivial bounds for the entanglement of the unconditional state at all times. 

These bounds display a richer dynamics than the exponential bounds previously considered, and closely follow the overall behavior of the exact dynamics (see Appendix~\ref{app:Exact}). The dependence of $C_{u^{\pm}}$ on the initial state is a generic feature of entanglement dynamics in systems in which more than one subsystem is coupled to a noisy channel, reflecting the varying effects of decoherence on different classes of states. In particular, for initial pure states which are not of \emph{full} dimension on the system Hilbert space, i.e., for which one or more of its components $\psi_{ij}$ vanish, $C_0 = X_0$ holds and the dynamics simplifies, resulting in the bounds $C_{u^{+}}=C_0 e^{-\gamma t}$ for concurrence and $C_{u^{+}}=C_0$ for concurrence of assistance,  which match the exact dynamics of the unconditional state entanglement. Thus for these states, which among others include all of Bell's maximally entangled states, the concurrence vanishes only asymptotically in time, while the concurrence of assistance remains constant and equal to the concurrence of the initial state at all times. These behaviors, however, are not generic for all initial states, as the long time limit of $C_{u^{\pm}}$ reveals. For general initial states $C_0 \le X_0$ holds, and hence \eqref{eq:Cpmdph} predicts the existence of a separation time $t_s < \infty$ after which $C_{u^{+}}(t\ge t_s)=0$ and the entanglement in the systems has disappeared, indicating that $\rho(t\ge t_s)$ is a separable state. The lower bound $C_{u^{-}}$ for the concurrence of assistance displays yet a different evolution: It increases monotonically to its asymptotic value $C_{u^{-}}(t\to\infty) = \frac{1}{2}(C_0 + X_0)$, which for a generic initial state indicates the generation of entanglement signaled by the growth of its concurrence of assistance with time.

\begin{figure}
	\includegraphics*[scale=0.6, bb= 45 45 450 341]{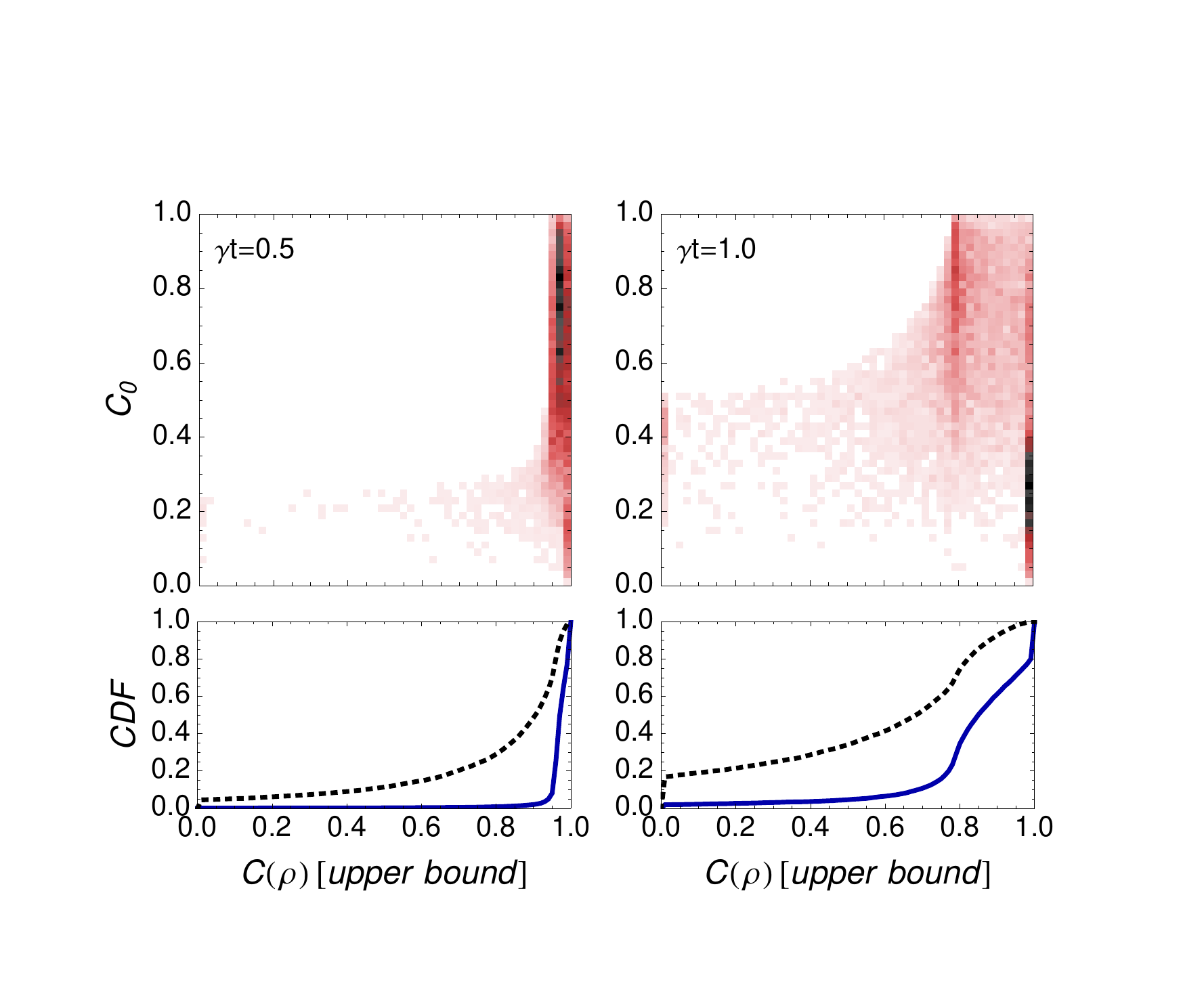}
	\caption{\label{fig:Cdph}Concurrence dynamical upper bound performance. The entanglement evolution of an ensemble of $10000$ initially pure states of a $2\times 2$ system coupled to phase damping channels is depicted for two different times. In the upper panels, \emph{points} indicate the concurrence of the final state $C(\rho)$ in units of the upper bound $C_{u^{+}}$ versus the initial state concurrence $C_0$. Darker regions contain more datapoints. The lower panels display the cumulative density function (CDF) (blue continuous line) corresponding to the probability density function of the upper panels. The CDF corresponding to the exponential bound (black dashed line) is also shown for comparison.}
\end{figure}
\begin{figure}
	\includegraphics*[scale=0.6, bb= 45 45 450 341]{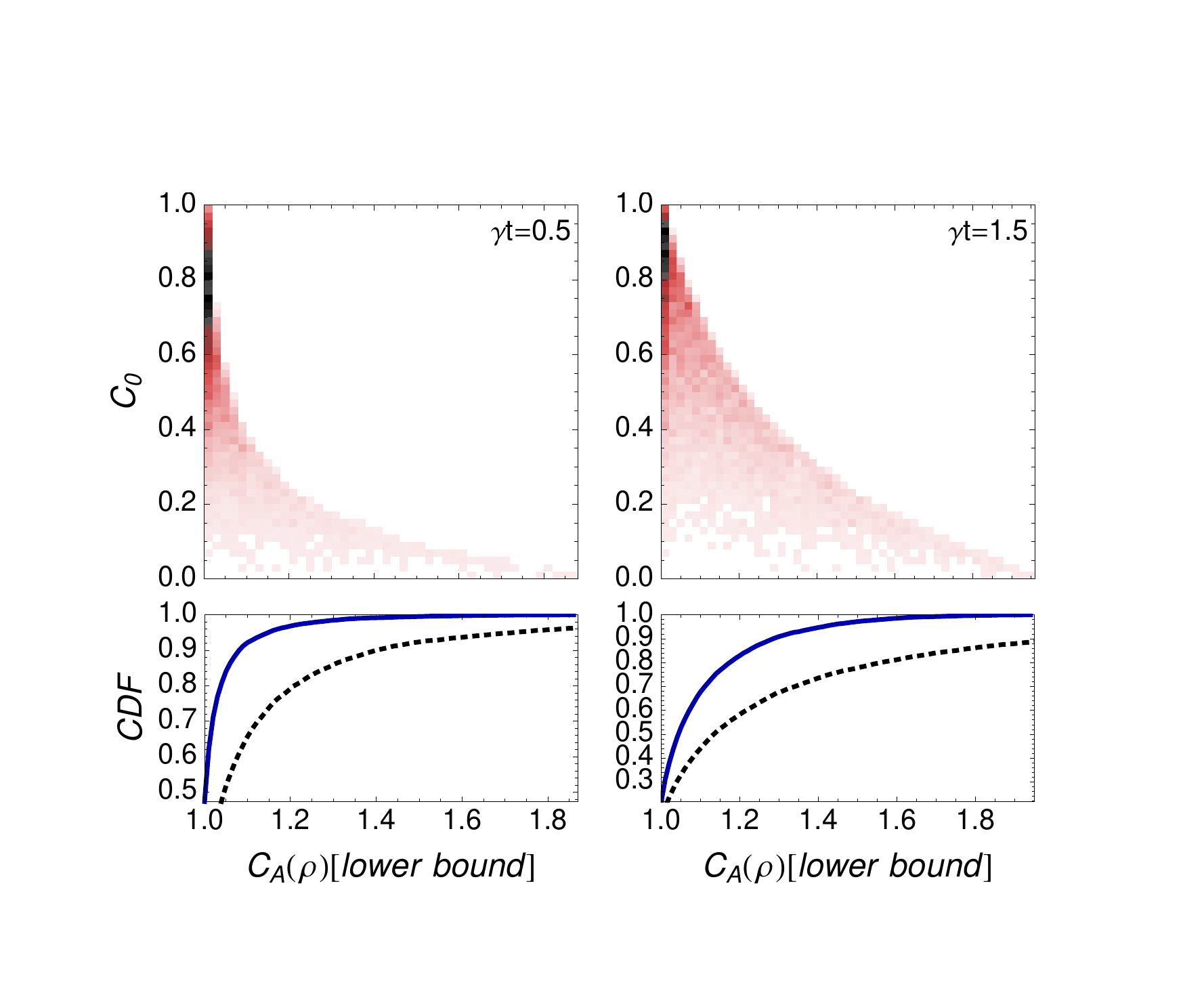}
	\caption{\label{fig:CAdph}Concurrence of assistance dynamical lower bound performance. The entanglement evolution of an ensemble of $10000$ initially pure states of a $2\times 2$ system coupled to phase damping channels is depicted for two different times. In the upper panels, \emph{points} indicate the concurrence of assistance of the final state $C_{A}(\rho)$ in units of the lower bound $C_{u^{-}}$ versus the initial state concurrence $C_0$. Darker regions contain more datapoints. The lower panels display the cumulative density function (CDF) (blue continuous line) corresponding to the probability density function of the upper panels. The CDF corresponding to the constant---entanglement protecting---bound (black dashed line) is also shown for comparison.}
\end{figure}

The overall performance of bounds $C_{u^{\pm}}$ approximating the exact concurrence $C(\rho(t))$ and concurrence of assistance $C_A(\rho(t))$ of the unconditional system state can be inferred from figures \ref{fig:Cdph} and \ref{fig:CAdph}, respectively. Since the obtained bounds can be evaluated for any initial pure state, we follow the entanglement dynamics of an ensemble of random uniformly distributed pure states on the Hilbert space \cite{Zycz01b}, avoiding in this way any bias on the numerical sampling. Figure \ref{fig:Cdph} shows that the upper bound gives a very good approximation for all initial states at all times. For highly entangled states this is particularly true for short times, as the density of points in the upper planes indicates. The capability of the bound to detect the separability of states at finite times guarantees its good performance also for larger times, as it is manifest in the right panels, where a concentration of points for initially low entangled states that have already became separable is clearly seen. These conclusions are quantitatively supported by the cumulative distribution function (CDF) displayed in the lower panels (blue continuous line), in which it can be recognized that even at large times the error for almost all states remains below $20\%$. Results for the exponential bound were discussed in Ref.~\cite{Tier09} and are shown (black dashed line) for comparison. The good quality of the approximation to concurrence of assistance provided by our lower bound is illustrated in Fig.~\ref{fig:CAdph}, where especially the behavior of highly entangled states is well reproduced. Its accuracy diminishes slowly with time as the entanglement in the system increases to eventually reach its long time asymptotic value, as can be quantitatively seen in the lower panels. 

\subsection{\label{Tch} Thermal bath}

In this section we consider a different two-qubit system, in which each of its qubits coupled to its own, independent thermal bath exchanging excitations with the environment. The Lindblad operators accounting for the action of the baths on the system are
\ben
\mathbf{J}= \left( \sqrt{\gamma_{-}}\sigma_{-}\otimes \mathbb{I}, \mathbb{I} \otimes \sqrt{\gamma_{-}}\sigma_{-}, \sqrt{\gamma_{+}} \sigma_{+}\otimes \mathbb{I}, \mathbb{I} \otimes \sqrt{\gamma_{+}}\sigma_{+} \right)^{\mathsf{T}},
\een
with emission  $\gamma_{-} = \gamma (\bar{n} + 1)$ and absorption $\gamma_{+} = \gamma \bar{n}$ rates, where $\gamma$ is the coupling amplitude to the baths and $\bar{n}$ is the mean number of excitations in the thermal baths, which we assume are kept at the same temperature. A $4\times 4$ matrix $u$ specifies the unravelings in this system.

The time evolution equation of the system average concurrence is found after explicit evaluation of Eq.~\eqref{eq:dcu},
\be
\label{eq:dCth}
\begin{split}
\frac{dC_{u}}{dt} &= - \left(\gamma_{-} + \gamma_{+} + \sqrt{\gamma_{-}\gamma_{+}}(u_{13}^{*} + u_{24}^{*})\right) C_{u} \\
& \quad + 2 \text{Re} \left( E\left[ \frac{c(\psi_c)}{C(\psi_c)} ( \gamma_{-}\psi_{c11}^{2} u_{12}^{*} + \gamma_{+}\psi_{c00}^{2} u_{34}^{*} \right. \right. \\
& \quad - \sqrt{\gamma_{-}\gamma_{+}} (\psi_{c10}^{2} u_{14}^{*} + \psi_{c01}^{2} u_{23}^{*} )  \biggr] \biggr).
\end{split}
\ee
In the following we consider the dynamics of $C_u$ in two limiting systems.

\subsubsection{\label{sec:T0}Zero temperature bath}

In the zero temperature limit $\bar{n}=0$ and thus the environment becomes an amplitude damping channel into which excitations in the system decay with rate $\gamma_{-}=\gamma$ while $\gamma_{+}=0$. Unravelings of the system are now parameterized by the left upper part of $u$ only. The equation of motion \eqref{eq:dCth} for the average concurrence reduces to
\be
\label{eq:dCT0}
\frac{dC_{u}}{dt} = - \gamma C_{u} + 2 \gamma \text{Re} \left( E\left[ \frac{c(\psi_c)}{C(\psi_c)}\psi_{c11}^{2} u_{12}^{*} \right] \right)\,,
\ee
which depends on the unraveling only through the non local correlations $u_{12}^{*}$.

Local monitoring of the environments implies $u_{12}=0$, and leads to an exponential decay of $C_u$, yielding the bounds
\be
\label{eq:CExpT0}
C(\rho(t)) \le C_{u}(t) = C_{0}e^{-\gamma t} \le C_{\text{A}}(\rho(t))
\ee
for the concurrence and concurrence of assistance of the unconditional system state.

The dynamical evolution of $C_u$ can nevertheless be maximized or minimized if non local measurements are performed. Simple inspection of Eq.~\eqref{eq:dCT0} shows that the changed in the mean concurrence is extreme for the choice 
\begin{subequations}
\begin{align}
u_{11}^{\pm} &= u_{22}^{\pm} = 0 \,, \\
u_{12}^{\pm} &= \mp e^{i\theta_{\text{opt}}}, 
\end{align}
\end{subequations}
of unraveling, with $\theta_{\text{opt}} = \arg (c(\psi_c)\psi_{c11}^{2})$. For these unravelings the equation of motion reads
\be
\frac{dC_{u^{\pm}}}{dt} = - \gamma C_{u} \mp 2 \gamma E[|\psi_{c11}|^{2}]\,,
\ee
which, after noticing that on average the population in state $\ket{11}$ is exponentially damped, $E[|\psi_{c11}|^{2}] = |\psi_{11}(0)|^{2}e^{-2\gamma t}$, reduces to the deterministic equation for the average concurrence
\be
\frac{dC_{u^{\pm}}}{dt} = - \gamma C_{u} \mp 2 \gamma |\psi_{11}(0)|^{2} e^{-2\gamma t},
\ee
with solutions that provide the bounds \cite{Vivi10,Masc11}
\be
\label{eq:dcT0}
C_{u^{\pm}}(t,\psi_0) = \max \left\{0, e^{-\gamma t} \left( C_0  \mp 2 |\psi_{11}(0)|^{2} (1-e^{-\gamma t}) \right) \right\} \,,
\ee
for the unconditional entanglement dynamics.

The striking  scenario of unraveling $u^{+}$ was thorough studied in \cite{Vivi10}, where it was proved that this upper bound coincides with the concurrence of the unconditional state $C(\rho(t))=C_{u^{+}}(t)$, and therefore presents an accurate description of the entanglement dynamics in the system. Here, we turn on the analysis of the lower bound for concurrence of assistance dispensed by unraveling $u^{-}$.

The lower bound $C_{u^{-}}$ reproduces all the qualitative features of the concurrence of assistance. In particular, it also encompasses the hallmark of the entanglement dynamics in this system, that is, its behavior clearly distinguish between two classes of initial states (see Appendix~\ref{app:Exact} for exact expressions for $C_{A}(\rho(t))$). For initial states for which $C_0 < 2|\psi_{11}(0)|$, $C_{u^{-}}$ increases for a time until it reaches a maximum, after which it exponentially decays to zero. The peculiar increase of concurrence of assistance for short times is also observed for some initial separable states belonging to this class of states, signaling the creation of entanglement within this unraveling. For all other states $C_{u^{-}}$ simply follows an exponential decay.

\begin{figure}
	\includegraphics*[scale=0.6, bb= 45 45 450 341]{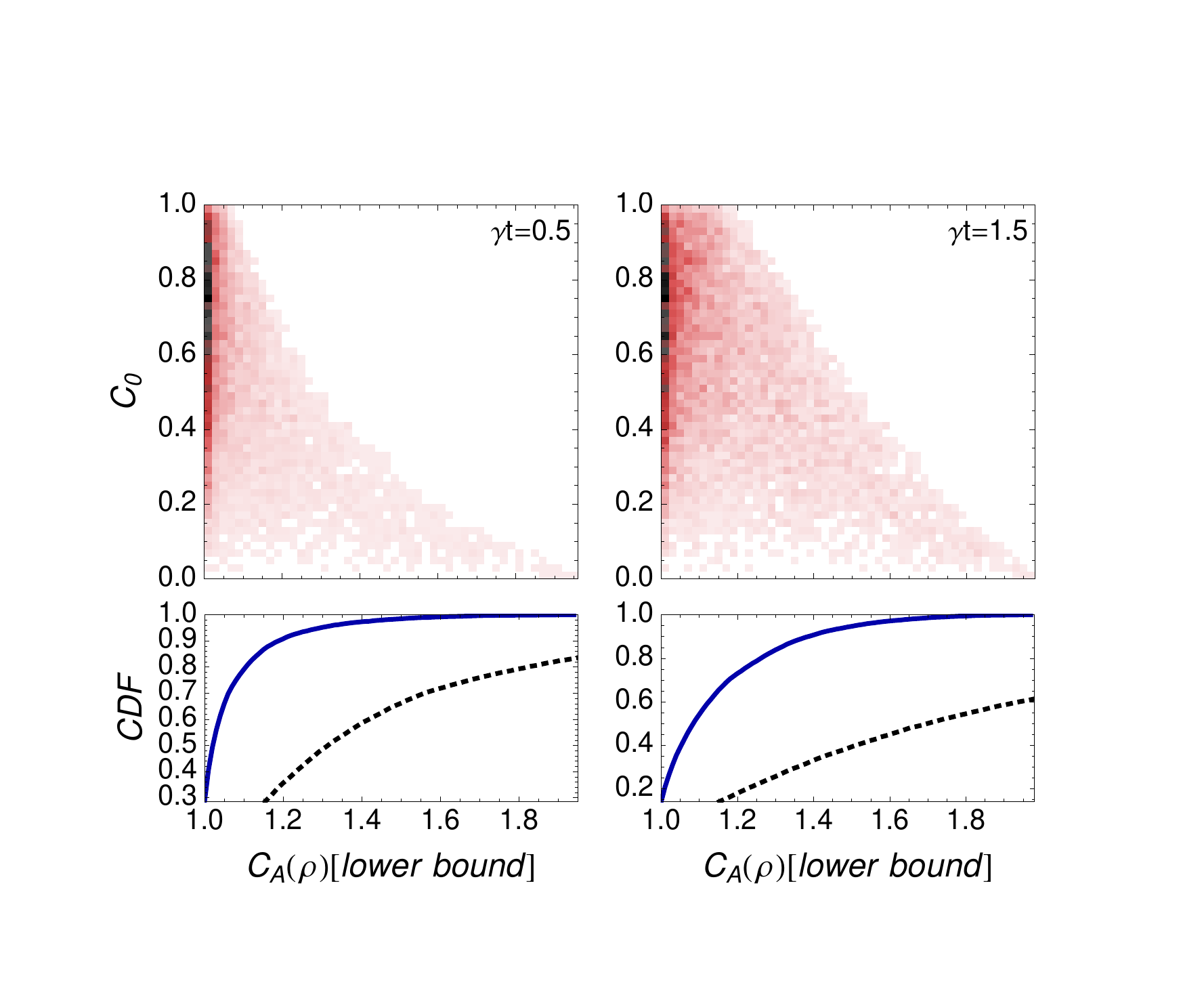}
	\caption{\label{fig:CAT0}Concurrence of assistance dynamical lower bound performance. The entanglement evolution of an ensemble of $10000$ initially pure states of a $2\times 2$ system coupled to amplitude damping baths is depicted for two different times. In the upper panels the \emph{points} indicate the concurrence of assistance of the final state $C_{A}(\rho)$ in units of the lower bound $C_{u^{-}}$ versus the initial state concurrence $C_0$. Darker regions contain more datapoints. The lower panels display the cumulative density function (CDF) (blue continuous line) corresponding to the probability density function of the upper panels. The CDF corresponding to the exponential bound in \eqref{eq:CExpT0} (black dashed line) is also shown for comparison.}
\end{figure}
That $C_{u^{-}}$ approximates $C_A(\rho(t))$ well in this system can be assessed with the help of Fig.~\ref{fig:CAT0}, in which the entanglement evolution for a uniform distribution of initial pure states is depicted. The lower bound provides a very good approximation for highly entangled states and short times. Even though for times of the order of the system characteristic time $1/\gamma$ the distribution of points in the upper panels becomes wider, for the large majority of states the lower bound (blue continuous line) underestimate the concurrence of assistance with an error smaller than the 20\%, as shown in the right lower panel. This good agreement is due to the aforementioned qualitative correspondence between $C_{u^{-}}$ and $C_A$, and can be  contrasted with the results obtained using a featureless bound as the exponential bound \eqref{eq:CExpT0} (black dashed line), which fails to capture the particularities of the concurrence of assistance evolution.

\subsubsection{\label{sec:Tinf}Infinite temperature}

In contrast to the previous section, where  we used our quantum trajectory formalism to build a picture of the entanglement dynamics of a quantum open system for all initial pure states, in this section we approach the study of the concurrence evolution for highly entangled states. This is a significant physical situation, since commonly implemented protocols in quantum information relay on highly entangled pure states for their success \cite{Ma:2013ei}. As it proves to be the case, for this kind of states our stochastic method offers analytical exact results for the system entanglement evolution.  

In a thermal bath at infinite temperature excitations and deexcitations in the system take place at the same rate $\Gamma$. This limit is reached by simultaneously taking the $\bar{n}\to\infty$ and $\gamma\to 0$ with $\gamma \bar{n} = \Gamma <\infty$. Equation~\eqref{eq:dCth} takes then the simpler form
\be
\label{eq:dCTinf}
\begin{split}
\frac{dC_{u}}{dt} &= - \Gamma \left(2 + (u_{13}^{*} + u_{24}^{*})\right) C_{u} \\
& \quad + 2 \Gamma\text{Re} \biggl( E\biggl[ \frac{c(\psi_c)}{C(\psi_c)} (\psi_{c11}^{2} u_{12}^{*} + \psi_{c00}^{2} u_{34}^{*} \\
& \quad - \psi_{c10}^{2} u_{14}^{*} - \psi_{c01}^{2} u_{23}^{*})\biggr] \biggr). 
\end{split}
\ee

Now, if only local measurements on the environments are allowed possible unravelings are restricted by the constrains $u_{12}=u_{14}=u_{23}=u_{34}=0$, and the average concurrence evolves exponentially in time with rate (cf. Eq.~\eqref{eq:CuExp})
\ben
k(u) = \Gamma \, \text{Re}(2 + u_{13}^{*} + u_{24}^{*})\,.
\een
It is then apparent that with a local monitoring strategy the best bound for the system unconditional state concurrence (concurrence of assistance) is obtained by choosing the unraveling with $u_{ii}=0$ for $i=1,2,3,4$, and $u_{13}=u_{24}=1$ ($u_{13}=u_{24}=-1$), leading to
\ben
C(\rho(t)) \le C_{0} e^{-4\Gamma t} \quad \text{ and } \quad C_A(\rho(t)) \ge C_{0}.
\een
As in the case of dephasing channels before, local measurements of the environment offer an exponential upper bound for the system concurrence and a constant lower bound, with a protecting unraveling, for the concurrence of assistance.

We now focus on the entanglement dynamics of maximally entangled states and restrict our initial states to Bell states $\ket{\Phi^{\pm}} = \frac{1}{\sqrt{2}}(\ket{00} \pm \ket{11})$ (equivalent results are obtained for states $\ket{\Psi^{\pm}} = \frac{1}{\sqrt{2}}(\ket{01} \pm \ket{10})$). Evidently, for completely entangled states the lower bound for the concurrence of assistance offered above is exact, $C_A(t)=1$. Hence we proceed to the concurrence dynamics for which better bounds are found if non local unravelings are considered. For the specified initial states in Appendix~\ref{app:Aux} we show that the time evolution of the upper bound $C_{u^{+}}$ is given by
\be
\label{eq:dCpTh}
\frac{dC_{u^{+}}}{dt} = -2\Gamma C_{u^{+}} - \Gamma (1 + e^{-4\Gamma t}),
\ee
which can be integrated without difficulty to obtain
\ben
C_{u^{+}}(t) = e^{-2 \gamma t}(C_0 - \sinh (2\gamma t)),
\een
showing that the state becomes separable at a finite time. 

The above result is noteworthy not only because it coincides with the known exact concurrence  for the specific physical scenario under study here \cite{Carv06}, but also because it was obtained without solving first the master equation for the unconditional dynamics. Although to arrive to equation \eqref{eq:dCpTh} it was necessary to integrate the time evolution of the populations on the system, this task demands a substantially smaller effort than the solution of the whole unconditional dynamics. The observation that the entanglement evolution of maximally entangled initial states can be accurately estimate within the quantum trajectory theory has also been reported for various other two-qubit systems \cite{Carv07,Vivi10,Voge10}.

\section{\label{sec:EoF} Entanglement of formation and localized unravelings}

Since our ideas stem from the stochastic evolution of entanglement along single quantum trajectories, the application of our method employing other entanglement measures for pure states proceeds smoothly. To illustrate this, we now use our scheme to address the relation between $C_u$ and the similarly defined average entanglement of formation 
\ben
EoF_{u}(t) \equiv E[EoF(\psi_c(t))],
\een
where the entanglement of the conditional state is obtained after evaluation of its von Neumann entropy $EoF(\psi_c) = -\tr(\rho_{c}^{(1)}\log_2 \rho_{c}^{(1)}) = -\tr(\rho_{c}^{(2)}\log_2 \rho_{c}^{(2)})$, with $\rho_{c}^{(1)}=\tr_{(2)}(\prj{\psi_c})$ and $\rho_{c}^{(2)}=\tr_{(1)}(\prj{\psi_c})$ given by the partial trace over one of the subsystems.

For two-qubit systems, however, entanglement of formation can be written as a function of concurrence \cite{Woot98},
\be
\label{eq:EoFC}
EoF(\psi_c) = h\left[\frac{1}{2}\left(1+\sqrt{1-C(\psi_c)^2}\right)\right]
\ee
with function $h(x) = -x \log_{2} x - (1-x)\log_{2}(1-x)$, which allow us to write its stochastic change along a quantum trajectory as function of the concurrence change,
\be
\label{eq:dEoF}
dEoF(\psi_c) =  \frac{dh}{dC}\biggr|_{\psi_c} dC(\psi_c) + \frac{d^2h}{dC^2}\biggr|_{\psi_c}   dC(\psi_c)^2 \,.
\ee
The It\^o form of the above equation of motion is reached after substitution of the explicit expressions for $dC(\psi_c)$ and its squared $dC(\psi_c)^2$. The ensemble average of the drift term determines the equation of motion for the average measure (see Appendix~\ref{app:E_cond_change} for details),
\be
\begin{split}
\label{eq:dEoFAvg}
\frac{d}{dt}EoF_u &= E\left[  \frac{dh}{dC}\biggr|_{\psi_c}  V(\psi_c,u) \right. \\
& \left. \quad +  2 \frac{d^2h}{dC^2}\biggr|_{\psi_c} \text{Re} \left( \mathbf{F}^{\mathsf{T}}(\psi_c) u^{*} \mathbf{F}(\psi_c) + |\mathbf{F}(\psi_c)|^2 \right) \right].
\end{split}
\ee
Due to its explicit dependence on the unraveling, the continuous measurements on the environment can be manipulated in order to generate ensembles of quantum trajectories exhibiting a desired entanglement property, which, in this case, may be better quantified by the entanglement of formation. 

While it is then possible to find unravelings that, for example, minimize the average of a particular entanglement measure but not of some other, it becomes of interest for the scheme proposed in this paper to be able to identify unravelings for which the attributes they imprint in the system entanglement, are independent of the entanglement measure. Specifically, for the case at hand and assuming local environments, we would like to determine under which conditions do unravelings $\bar{u}$ exist for which relation \eqref{eq:EoFC} can be extended to the average measures, that is,  
\be
\label{eq:EoFu}
EoF_{\bar{u}}(t)=h\left[\frac{1}{2}\left(1+\sqrt{1-C_{\bar{u}}(t)^2}\right)\right].
\ee

Observe that the above statement is true if $EoF_{\bar{u}}(t)$ is a solution of the equation of motion \eqref{eq:dEoFAvg}. Thus, after direct substitution, and comparing terms with derivatives of $h$ of the same order, we arrive at equations 
\begin{subequations}
\begin{gather} 
\label{eq:ubarC1}
\frac{dh}{dC}\biggr|_{C_{\bar{u}}}  E[V(\psi_c,\bar{u})] = E\left[  \frac{dh}{dC}\biggr|_{\psi_c}  V(\psi_c,\bar{u})\right], \\
\intertext{and}
\label{eq:ubarC2}
\mathbf{F}^{\mathsf{T}}(\psi_c) \bar{u}^{*} \mathbf{F}(\psi_c) + |\mathbf{F}(\psi_c)|^2 = 0,
\end{gather}
\end{subequations}
which provide the conditions to determine $\bar{u}$. Here, to write the left hand side of the first equation we used equation of motion \eqref{eq:dcu} to substitute the time derivative of the average concurrence, while the second equation results from demanding that the term proportional to the second derivative of $h$ in \eqref{eq:dEoFAvg} vanishes. As we now demonstrate, both equations are satisfied if the noise correlations are adaptively chosen as
\be
\label{eq:ubar}
\bar{u} = -\frac{F_{k}(\psi_c)}{|F_{k}(\psi_c)|} \delta_{kl}.
\ee
That this choice of $\bar{u}$ is a solution of \eqref{eq:ubarC2} is evident. To verify that it also provides a solution to \eqref{eq:ubarC1} demands a little more effort. Consider for a moment the stochastic evolution of concurrence along single trajectories given by \eqref{eq:dcon} when unraveling $\bar{u}$ is implemented. Since the environments are local, the diagonal form of $\bar{u}$ implies that the environment measurement setup is local too, and therefore the drift term factorizes $V(\psi_c,\bar{u})dt =-k(\bar{u})C(\psi_c)dt$. In addition, as can be verified by inspecting the correlation relations \eqref{eq:dxi}, the noises corresponding to $\bar{u}$ are given by $d\xi_k = i (F_{k}(\psi_c)/|F_{k}(\psi_c)|)^{1/2} dW$ with real Wiener increments satisfying $dW_k dW_l = \delta_{kl}dt$ and hence the noise term $\text{Re}\left[d\boldsymbol{\xi}^{\dagger}\mathbf{F}(\psi_c) \right] $ vanishes. As a result, equation of motion  \eqref{eq:dcon} reduces to $dC(\psi_c)/dt = -k(\bar{u})C(\psi_c)$, and the time evolution of concurrence along \emph{single} trajectories of unraveling $\bar{u}$ is no longer random but deterministic, and thus the same for \emph{all} trajectories in the ensemble. Consequently, the average concurrence is trivially computed to be $C_{\bar{u}}(t) = C(\psi_c)$, and in general for any function of concurrence  $f(C(\psi_c)) = f(C_{\bar{u}})$ holds. In particular, for the derivative of $h$ respect to concurrence we obtain $dh/dC|_{\psi_c} = dh/dC|_{C_{\bar{u}}}$, showing that \eqref{eq:ubar} is a solution of \eqref{eq:ubarC1} too. 

Our findings are properly illustrated in the following example. In a $2 \times 2$ system coupled to amplitude damping channels and initially prepared in the maximally entangled state $\ket{\Psi_{\pm}}=\frac{1}{\sqrt{2}}(\ket{01} \pm \ket{10})$, the evolution of the  average concurrence is given by $C_{u}(t) = e^{-\gamma t}$, which not only is independent of the unraveling, but also coincide with the unmonitored concurrence evolution (cf. Eq.~\eqref{eq:dcT0}). That is, for this particular physical setup, \emph{all} possible ways in which the environment can be continuously measured yield the same value for the average concurrence in the system, yet on the distinct ensembles of trajectories generated the fluctuations of entanglement differ, as can be attested if instead of concurrence one uses entanglement of formation as entanglement measure. Indeed, the \emph{only} unraveling for which the average entanglement of formation exactly reproduces the evolution of the entanglement of formation of the unconditional state $\rho$, and \eqref{eq:EoFu} is satisfied, is $\bar{u}$. For all other unravelings one finds that the average entanglement of formation provides an upper bound to the unconditional entanglement evolution $EoF(\rho(t))\le EoF_{u}(t)$.

More significantly, however, than the proof of relation \eqref{eq:EoFu} is that along the way we have shown a unique feature of unraveling $\bar{u}$: on its ensemble of trajectories concurrence does not fluctuate, but localizes along its mean value with a zero width distribution. This observation, based on the analysis of equation \eqref{eq:dcon}, goes beyond concurrence or entanglement of formation and constitute a general statement about entanglement in the system independently of the entanglement measure. Notice, for example, that localization was an essential property of the entanglement protecting protocol proposed in Section~\ref{sec:local}, and is therefore independent of the entanglement measure. The existence of such localized unravelings give way to interesting applications of our method, both experimentally an theoretically, as all the information about the entanglement evolution in the system is encoded in a \emph{single} trajectory and therefore single realizations of the system are sufficient for a comprehensive analysis of it \cite{Vivi10,Vivi14b}.  Besides, the adaptability of the method to different entanglement measures complements its already established adjustability regarding the choices of unravelings enhancing its overall strength as a tool for the characterization of entanglement time evolution.

\section{Summary}

In this work we have presented a thorough implementation of the quantum trajectory theory for the description of the entanglement time evolution in a Markovian open quantum system made of two qubits. To characterize the entanglement in the ensemble of trajectories unraveling the system dynamics we introduced the average concurrence and derived for it a deterministic equation of motion, providing in this way a comprehensive description of the entanglement evolution in the system. Remarkably, this complete picture is achieved without having to specified the state of the system beyond its initial configuration, i.e., for times larger than the initial time, conferring the method an efficiency that the usual approaches to the study of the entanglement dynamics in open systems lack. The most significant contribution of our proposal, however, is its versatility, which stems from essential dynamical consideration of the quantum trajectory formalism: Different measurement schemes use to monitor the environment account for different unravelings of the system dynamics and, consequently, generate ensembles of trajectories with distinct statistical properties. In particular, the average concurrence in the system depends on the ways the environment is being continuously monitored and therefore, in this sense, can be controlled. We exploited this flexibility to address two different issues of relevance in quantum information with our method: entanglement protection and entanglement estimation.

Regarding the first issue, for open two-qubit systems in which the effects of the environments are described by local, hermitian Lindblad operators we identified the existence of a local unraveling leading to a perfect protection of the entanglement in the system. Notably, for this protecting unraveling not only the average entanglement does not change in time, remaining equal to its initial value, but it does so because the entanglement is protected on a single trajectory basis despite the stochastic evolution of the state along it. As for the entanglement estimation, we demonstrated the capability of our method to provide analytical tight bounds for the concurrence and concurrence of assistance for the unmonitored dynamics of the system in various relevant cases, including coupling to dephasing and thermal noisy channels. Our bounds work for all times and are efficiently found without having to solve the unmonitored dynamics for the system state. Finally, we have also discussed the independence of our results on the choice of entanglement measure, and in the course of it showed the existence of localized unravelings, that is, unravelings for which along single trajectories the entanglement evolves smoothly in time despite the stochastic evolution of the system conditional state. Strikingly, in these cases, a single trajectory offers a complete description of the entanglement dynamics in the system \cite{Vivi10}.

To conclude it is worth stressing that our method is not restricted by the small size of the system we used to illustrate them, nor by our choice of concurrence as entanglement measure. Along the whole paper we had made a consciously effort to clearly lay the ground work that allow the extension of our results to more general physical situations, comprising systems of larger dimensions and number of parties, as well as different entanglement measures \cite{Vivi14b}. In the case of entanglement protection, non efficient detection is a relevant issue that must be addressed in future works.

\begin{acknowledgments}
We kindly acknowledge illuminating discussions with Juan Diego Urbina. C.V. is thankful for the hospitality extended to him by Andreas Buchleitner at the Physikalisches Institute Albert-Ludwigs Universit\"at Freiburg and Tobias Brandes at the Institut f\"ur Theoretische Physik at the Technische Universit\"at Berlin where the last part of this work was done. Financial support by Fundaci\'on para la Promoci\'on de la Investigaci\'on y la Tecnolog\'{\i}a (Colombia) under project No.2765 and by Divisi\'on de Investigaci\'on -- Universidad Nacional de Colombia under project No.10870  is gratefully acknowledged.
\end{acknowledgments}

\appendix

\section{\label{app:E_cond_change} Entanglement increment}

\subsection{Concurrence increment}

In this appendix we calculate the concurrence increment along a quantum trajectory given by \eqref{eq:dcon}. We start by noticing that for any complex function of the system conditional state, $g=g(\psi_{c})$, the It\^{o} form of the change of its norm can be written in terms of the change $dg^{*}$ as 
\be
\label{eq:dAbs}
d|g| = \frac{1}{|g|}\text{Re}\,(g\, dg^{*})+\frac{1}{2|g|^{3}}\left[\text{Im}(g\,dg^{*})\right]^{2}\,.
\ee

In particular, we consider the preconcurrence increment $dc(\psi_{c})^{*} = d\ovl{\tilde{\psi}_{c}}{\psi_{c}} = \ovl{d\tilde{\psi}_{c}}{\psi_{c}} + \ovl{\tilde{\psi}_{c}}{d\psi_{c}} + \ovl{d\tilde{\psi}_{c}}{d\psi_{c}}$ which is easily evaluated by means of the evolution equation \eqref{eq:diffeq} for the conditional state $d\ket{\psi_{c}}$,
\be
\label{eq:dc}
dc(\psi_{c})^{*} = \left(2 \ovl{\tilde{\psi_{c}}}{v} + \sum_{k,l} \ovl{\tilde{f_k}}{f_l}u^*_{kl}\right)dt + 2 d\boldsymbol{\xi}^{\dagger}\ovl{\tilde{\psi_{c}}}{\mathbf{f}}\,.
\ee 
To obtain the change in the concurrence we now use relation \eqref{eq:dAbs} together with \eqref{eq:dc}, which, after some straight forward simplifications and the use of the noise properties \eqref{eq:dxi}, yield
\begin{equation*}
\begin{split}
dC(\psi_c) &= \text{Re}\left[\frac{c(\psi_c)}{C(\psi_c)} \left( 2\ovl{\tilde{\psi_c}}{v} + \frac{1}{c(\psi_c) }|\ovl{\tilde{\psi}_c}{\mathbf{f}}|^2 \right.\right.\\
& \quad \left. \left.- \frac{c(\psi_c)}{C(\psi_c)^2} \ovl{\tilde{\psi_c}}{\mathbf{f}^{\mathsf{T}}} u^* \ovl{\tilde{\psi}_c}{\mathbf{f}} + \sum_{kl}\ovl{\tilde{f_k}}{f_l} u_{kl}^*\right)dt \right.\\
& \quad \left.+ 2 d\boldsymbol{\xi}^{\dagger} \ovl{\tilde{\psi}}{\mathbf{f}}\right] \,.
\end{split}
\end{equation*}
We arrive to the desired equation~\eqref{eq:dcon} after substituting in the expression above the explicit forms \eqref{eq:drift} and \eqref{eq:diff} for the drift and noise amplitudes of the state change, respectively. 

Entanglement dynamical equation \eqref{eq:dcon} simplifies for local Lindblad operators. In this case  the following relations hold \cite{Voge10},
\begin{align*}
\avg{\widetilde{J_k}}{\psi} & = \frac{1}{2}c(\psi)^*(\tr_{\mathbb{C}^2} J_k), \\
\ovl{\widetilde{J_k \, \psi}}{J_l \, \psi} &= -\frac{1}{2}c(\psi)^* \left[\tr_{\mathbb{C}^2}(J_k J_l) - \tr_{\mathbb{C}^2}(J_k) \tr_{\mathbb{C}^2}(J_l) \right],
\end{align*}
for operators $J_k$ and $J_l$ acting on the same qubit, and the trace is taken over a single qubit space. If in addition we impossed that only local measurements are done, i.e.,  $u_{kl} = 0$ if $J_k$ and $J_l$ are operators acting on different qubits, the deterministic amplitude \eqref{eq:V} can be recast as a product of a factor and the system conditional concurrence,
\ben
V(\psi_{c},u) = - k(u)C(\psi_{c})\,,
\een
where $k(u)$, given by Eq.~\eqref{eq:ku}, contains all the information about the unraveling. Similarly, the noise amplitude reduces to
\ben
\mathbf{F}(\psi_c) =\left(\frac{1}{2}\tr_{\mathbb{C}^2}\mathbf{J}  -\avg{\mathbf{J}}{c} \right)  C(\psi_c),
\een
becoming proportional to the concurrence too.
 
\subsection{Entanglement of formation increment} 

The It\^o form of the entanglement of formation change along a quantum trajectory demands the evaluation of $dC(\psi_c)^2$. Expression~\eqref{eq:dcon} for $dC(\psi_c)$ permits to do this fast. Making use of the noise correlations \eqref{eq:dxi} and keeping terms up to first order in $dt$, we obtain
\ben
\label{eq:dCdC}
(dC_{\psi_c})^2 = 2 \text{Re} \left( \mathbf{F}^{\mathsf{T}} u^{*} \mathbf{F} + |\mathbf{F}|^2 \right)dt .
\een
The sought equation of motion is reached after substitution of the explicit expressions \eqref{eq:dcon} for $dC(\psi_c)$, and the above expression for its squared, into \eqref{eq:dEoF} 
\be
\label{eq:dEoFIto}
\begin{split}
dEoF(\psi_c) &= \left[ \frac{dh}{dC}\biggr|_{\psi_c}  V(\psi_c,u) \right. \\
& \quad \left. + 2 \frac{d^2h}{dC^2}\biggr|_{\psi_c} \text{Re} \left( \mathbf{F}^{\mathsf{T}}(\psi_c) u^{*} \mathbf{F}(\psi_c) + |\mathbf{F}(\psi_c)|^2 \right) \right] dt\\
& \quad + 2  \frac{dh}{dC}\biggr|_{\psi_c}  \text{Re}(d\boldsymbol{\xi}^{\dagger}\mathbf{F}(\psi_c)) \,.
\end{split}
\ee
 
\section{\label{app:Aux} Stochastic increments for auxiliary quantities}

\subsection{Dephasing channel}

Here we derive the equation of motion for the ensemble averaged function $E[ X_{\psi} ] = 2 E[|(\psi_{01}\psi_{10})| + |(\psi_{00}\psi_{11})|]$ in the system of two-qubit described in Section~\ref{sec:Dch}. 

The stochastic changes $d|(\psi_{01}\psi_{10})|$ and $d|(\psi_{00}\psi_{11})|$ are evaluated using relation \eqref{eq:dAbs}. The needed increments $d(\psi_{01}\psi_{10}) =  \ovl{01}{d\psi} \ovl{10}{\psi} +  \ovl{01}{\psi} \ovl{10}{d\psi}+  \ovl{01}{d\psi} \ovl{10}{d\psi}$ and $d(\psi_{00}\psi_{11}) = \ovl{00}{d\psi} \ovl{11}{\psi} +  \ovl{00}{\psi} \ovl{11}{d\psi} +  \ovl{00}{d\psi} \ovl{11}{d\psi}$ are obtained after explicit substitution of the state increment \eqref{eq:diffeq},
\begin{align*}
\begin{split}
d|(\psi_{01}\psi_{10})| &= -\frac{\gamma}{4} |(\psi_{01}\psi_{10})| \left[ 2 - \text{Re}(u_{11}^{*} + u_{22}^{*} - 2 u_{12}^{*})\right] \\
& \quad + \sqrt{\gamma}\left[ 1-2\left(|\psi_{00}|^2-|\psi_{01}|^2\right) \right] d\xi_{1}^{*} \\
& \quad + \sqrt{\gamma}\left[ 1-2\left(|\psi_{00}|^2-|\psi_{10}|^2\right) \right] d\xi_{2}^{*} \,,
\end{split} \\
\begin{split}
d|(\psi_{00}\psi_{11})| &= -\frac{\gamma}{4} |(\psi_{00}\psi_{11})| \left[ 2 - \text{Re}(u_{11}^{*} + u_{22}^{*} + 2 u_{12}^{*})\right] \\
& \quad + \sqrt{\gamma}\left[ 1-2\left(|\psi_{00}|^2-|\psi_{01}|^2\right) \right] d\xi_{1}^{*} \\
& \quad + \sqrt{\gamma}\left[ 1-2\left(|\psi_{00}|^2-|\psi_{10}|^2\right) \right] d\xi_{2}^{*}.
\end{split}
\end{align*}
Upon addition, and after evaluation of the ensemble average, the unraveling dependent equation o motion for $E[ X(\psi_c) ]$ is
\ben
\begin{split}
\frac{d}{dt} E[ X(\psi_c) ] &= -\frac{\gamma}{2} E[ X(\psi_c) ] \\
& \quad - \frac{\gamma}{2} \text{Re}\left\{ E[ |\psi_{c01}\psi_{c10}| (u_{11}^{*} + u_{22}^{*} - 2 u_{12}^{*}) \right. \\
& \quad \left. + |\psi_{c00}\psi_{c11}| (u_{11}^{*} + u_{22}^{*} + 2 u_{12}^{*})]\right\}\,.
\end{split}
\een

Equation~\eqref{eq:AD} follows after substitution of the choice \eqref{eq:Duopt} for the unraveling.

\subsection{\label{app:AuxTInf} Infinite temperature bath channel}

In this section we derive the equation of motion \eqref{eq:dCpTh} for $C_{u^{+}}$ starting from Eq.~\eqref{eq:dCTinf} and accounting for non local unravelings only. An inspection of \eqref{eq:dCTinf} shows that for a minimization of its right hand the matrix $u$ must be of the form
\ben
u = e^{i \theta_c}
\begin{pmatrix}
0 & - \alpha_{12}\, e^{2i \theta_{11}} & 0 & \alpha_{14}\, e^{2i \theta_{10}} \\
- \alpha_{12}\, e^{2i \theta_{11}}  & 0 & \alpha_{23}\, e^{2i \theta_{01}} & 0 \\
0 & \alpha_{23}\, e^{2i \theta_{01}} & 0 & - \alpha_{34}\, e^{2i \theta_{00}} \\ 
\alpha_{14}\, e^{2i \theta_{10}}  & 0 & -\alpha_{34}\, e^{2i \theta_{00}} & 0
\end{pmatrix}
\een
with $\theta_{11} = \arg (\psi_{c11})$, $\theta_{10} = \arg (\psi_{c10})$, $\theta_{01} = \arg (\psi_{c01})$, $\theta_{00} = \arg (\psi_{c00})$ and parameters $\alpha_{11}$, $\alpha_{10}$, $\alpha_{01}$, and $\alpha_{11}$ to be determined under the constrain that $u$ remains physical (cf. Eq.~\eqref{eq:R}). The equation of motion then simplifies to
\ben 
\begin{split}
\frac{dC_{u}}{dt} &= -2\Gamma C_u - 2 \Gamma\text{Re} \bigl( E\bigl[ \alpha_{11}|\psi_{c11}|^{2} + \alpha_{00}|\psi_{c00}|^{2} \\
& \quad + \alpha_{10}|\psi_{c10}|^{2} + \alpha_{01}|\psi_{c01}|^{2}\bigr] \bigr). 
\end{split}
\een

To complete the above dynamics we notice that the evolution of the populations are unraveling independent and their averages are straightaway to integrate. With the initial conditions given by the states $\ket{\Phi^{\pm}}$ they are
\begin{align*}
E[|\psi_{c11}|^{2}](t) &= E[|\psi_{c00}|^{2}](t) = \frac{1}{4}(1 + e^{-4\Gamma t}),\\
E[|\psi_{c10}|^{2}](t) &= E[|\psi_{c01}|^{2}](t) = \frac{1}{4}(1 - e^{-4\Gamma t}).
\end{align*}

After substitution, the equation of motion for the average concurrence now reads
\ben 
\begin{split}
\frac{dC_{u}}{dt} &= -2\Gamma C_u - \frac{1}{2} \Gamma\text{Re} \bigl( E\bigl[ (\alpha_{11}+\alpha_{00}) (1 + e^{-4\Gamma t}) \\
& \quad + (\alpha_{10}+\alpha_{01}) (1 - e^{-4\Gamma t})\bigr] \bigr). 
\end{split}
\een 
A minimizing dynamics is obtained with the choices $\alpha_{11}=\alpha_{00}=1$ and $\alpha_{10}=\alpha_{10}=0$, settling unraveling $u^{+}$ and yielding equation \eqref{eq:dCpTh} for the upper bound $C_{u^{+}}$. 	

\section{\label{app:Exact}Unconditional state concurrence and concurrence of assistance.}

In this section, for the purpose of reference, we list exact expressions for the evolution of concurrence and concurrence of assistance of the unconditional system state in the systems considered in Section~\ref{sec:RhoE} of the main text, valid when initial pure states are considered.

In a two-qubit system in which each subsystem couples independently to a dephasing channel the concurrence~\cite{Fons12} and concurrence of assistance are, respectively,
\begin{align*}
C(\rho(t)) &=\frac{1}{2}\biggl(-(1-e^{-\gamma t})X_0 \\
&\quad \left. + \sqrt{ (1-e^{-\gamma t})^2 W_0^2+ 4e^{-\gamma t} C_0^2 }\right), \\
C_A(\rho(t)) &= \frac{1}{2}\biggl((1-e^{-\gamma t})X_0 \\
&\quad \left.+ \sqrt{ (1-e^{-\gamma t})^2 X_0^2+ 4e^{-\gamma t} C_0^2 }\right),
\end{align*}
where we introduced the state function $W(\psi_c) =2( |\psi_{c01}\psi_{c10}| - |\psi_{c00}\psi_{c11}|)$ and $W_0=W(\psi(0))$.

The concurrence \cite{Vivi10}  and concurrence of assistance in a two-qubit system in which each subsystem couples independently to a zero temperature bath are
\begin{align*}
C(\rho(t)) &= e^{-\gamma t}\left[C_0 -2 |\psi_{11}(0)|^2 (1-e^{-\gamma t})\right], \\
C_A(\rho(t)) &= e^{-\gamma t}\left[2 |\psi_{11}(0)|^2 (1-e^{-\gamma t}) \right.\\
& \quad \left. + \sqrt{4 |\psi_{11}(0)|^4 (1-e^{-\gamma t})^2 + C_0^2}\right],
\end{align*}
respectively.

%
\end{document}